\documentclass[aps,twocolumn,showpacs,floatfix,superscriptaddress]{revtex4}
\usepackage{graphicx}
\usepackage{psfrag}
\begin{document}
\newcommand{\beq}{\begin{equation}}
\newcommand{\eeq}{\end{equation}}
\newcommand{\ben}{\begin{eqnarray}}
\newcommand{\een}{\end{eqnarray}}
\newcommand{\bea}{\begin{array}}
\newcommand{\eea}{\end{array}}
\newcommand{\om}{(\omega )}
\newcommand{\bef}{\begin{figure}}
\newcommand{\eef}{\end{figure}}
\newcommand{\leg}[1]{\caption{\protect\rm{\protect\footnotesize{#1}}}}
\newcommand{\ew}[1]{\langle{#1}\rangle}
\newcommand{\be}[1]{\mid\!{#1}\!\mid}
\newcommand{\no}{\nonumber}
\newcommand{\etal}{{\em et~al }}
\newcommand{\geff}{g_{\mbox{\it{\scriptsize{eff}}}}}
\newcommand{\da}[1]{{#1}^\dagger}
\newcommand{\cf}{{\it cf.\/}\ }
\newcommand{\ie}{{\it i.e.\/}\ }
\title{Magnetic Reversal Time in Open Long Range Systems}
\author{F.~Borgonovi}
\affiliation{Dipartimento di Matematica e Fisica, Universit\`a Cattolica, via Musei 41, 25121
Brescia, Italy}
\affiliation{I.N.F.N., Sezione di Pavia, Italy}
\author{G.~L.~Celardo}
\affiliation{Instituto de Fisica, Benemerita Universidad Autonoma de Puebla,
Puebla, Mexico}
\author{B.~Goncalves}
\affiliation{Emory Univ., Atlanta USA}
\author{L.~Spadafora}
\affiliation{Dipartimento di
Matematica e Fisica, Universit\`a Cattolica, via Musei 41, 25121
Brescia, Italy}
\begin{abstract}
Topological phase space disconnection  has been recently  
found to be a general 
phenomenon in isolated anisotropic spin systems.  
It sets a general framework to understand
the emergence of ferromagnetism in finite magnetic systems
starting from microscopic models without phenomenological on-site
barriers.
Here we study its relevance for finite systems with long range
interacting potential in contact with a thermal  bath.
We show that, even in this case,  
the induced magnetic reversal time is
exponentially large in the number of spins,
thus determining {\it stable} (to any experimental observation time) 
ferromagnetic behavior.
Moreover, the explicit temperature dependence of the magnetic reversal time
obtained from the microcanonical 
results, is found to be in good agreement with numerical simulations.
Also,  a simple and suggestive  expression, indicating
the Topological Energy Threshold at which the 
disconnection occurs,  as a real energy barrier for many body 
systems, 
is obtained analytically for  low temperature.
\end{abstract}
\date{\today}
\maketitle

\section{Introduction}

One dimensional toy 
models are widely investigated in statistical mechanics\cite{oned}.
From one hand, the possibility to get  analytical results represents 
the starting
point for analyzing more physical models. On the other hand, 
due to their high simplicity,
they allow a better understanding of the key mechanisms at the basis 
of important
physical effects.
It is the case of the Topological Non-connectivity Threshold (TNT), 
recently introduced and addressed
in \cite{jsp} and investigated  in other related papers 
\cite{firenze,tutti,ultimoruffo}.
In these simple toy models, with a well defined classical 
limit two key features were
introduced, anisotropy and long--range coupling. 
Even if acting in different ways,
they are both essential to generate a   significant disconnection
 of the Hamiltonian phase space   leading to  what is known
in literature as  breaking of
 ergodicity \cite{palmer, firenze}. While anisotropy is a common paradigm
 in the phenomenological models of ferromagnetism
(usually introduced as on-site anisotropy barrier in microscopic
models)
\cite{cenetanti},
 long range
 interactions were re-discovered quite recently, due to the development
  of powerful  and efficient techniques\cite{libroruffo}.
Indeed, strictly speaking, a well known model for anisotropy including
a $-M^2$ term in the magnetic energy\cite{chudnovsky} ($M=\sum S_i$
being the sum of all magnetic moments within a suitable domain) exactly
matches an all-to-all interacting model close to what we consider here
below. 
The role of anisotropy in finite spin systems has attracted 
much attention recently,
following the experimental verification of ferromagnetic behavior
in finite 1D systems with strong anisotropy
\cite{gambnat},
contrary to common expectation that ferromagnetic behavior is
proper of macroscopic systems only. 
Theoretical works 
\cite{gamb2,cinesi},
attempted to  explain such ferromagnetic behavior in finite 
systems using microscopic models
with on site anisotropy,  inducing  an effective energy
barrier and thus leading to large magnetic reversal times and
to ferromagnetic behavior due to finite measurement time.

In this paper we take a different  approach modeling both anisotropy 
and long-range with some suitable spin--spin interaction toy model
as in \cite{tutti} but, differently from there,
and in order to produce results closer to real experiments,
we put the system in contact with a thermal bath.
Despite its simplicity, it can be easily fitted to more physical
models: for instance
spin systems with dipole interaction in 3-D, have both long range
and anisotropic spin-spin interactions.

There are many different ways to model a thermal environment,
especially when thermalization of a long range system is needed. Here
we take the simplest route 
and assume that the environment is able
to produce
a Gibbs distribution for the system energies.

In \cite{firenze} we have found, for isolated anisotropic systems 
with an easy axis  of magnetization (defined
as the direction of the magnetization in the ground state
energy configuration)  that 
the constant energy surface  is disconnected in   two regions 
with positive and negative magnetization along the easy axis. 
This disconnection occurs below a critical energy 
threshold which has been called 
Topological Non-connectivity Threshold.  
Since the phase space is disconnected  we have exact 
ergodicity breaking and no
dynamical trajectory can  visit the whole
constant energy surface, but is, instead, 
limited to the region in which
it started. 
Moreover, being defined  for all finite $N$, 
where $N$ is the  number of spins, 
the  ergodicity breaking is not related to the thermodynamic limit.
The Topological Non-connectivity Threshold is an example of topological 
singularity, well  studied recently \cite{pettini},  
and its existence  has been  pointed out in
Ising models too\cite{ultimoruffo}. 
Also, an experimental test
of ergodicity breaking has been proposed
\cite{expramaz}.

Even if  the connection between ergodicity breaking and 
anisotropy 
has been shown to be a general one, independent of the range 
of interaction among the spins, long range interacting systems
behave quite differently from short ones in the thermodynamic limit.
This consequence has been studied in 
\cite{tutti}  
where it  was found that the disconnected  portion of the energy spectrum 
remains finite in the thermodynamic limit for long-range  interacting 
systems only, while it  becomes negligible, in the same limit, 
for short-range  interacting ones. 

The plan of the paper is the following: In Sec.~II, we review and extend 
the results obtained 
in Ref.~\cite{firenze, tutti} concerning the microcanonical behavior
of long range interacting systems. 
In Sec.~III we present a detailed calculation of the density
of states, using large deviation techniques, for a Mean Field 
Hamiltonian 
and  we compared it for a generic long range interacting system.
Finally in Sec.~IV we study the magnetic reversal time in the 
canonical ensemble.
In particular,  we show how the  magnetic reversal time for
the open system can be obtained from that of the isolated one 
(a non trivial result). Also, a very
simple approximation allows to interpret the $E_{tnt}$
as a real energy barrier for many body spin systems at sufficiently low
temperature.

\section{The TNT: microcanonical results}

Our paradigmatic anisotropic model, with an adjustable interaction range,
which presents ergodicity breaking for any $N$, is described by the 
following Hamiltonian:

\begin{equation}
\label{H}
H=  J\left( {\eta \over 2}\sum_{i\not=j}^N {S^x_i S^x_j \over r_{ij}^\alpha}
-      {1 \over 2}\sum_{i\not=j}^N {S^z_i S^z_j \over r_{ij}^\alpha}
\right),
\end{equation}
\noindent
where  
$S^x_i, S^y_i, S^z_i$ are the spin components, assumed to vary continuously; 
$i,j=1,...N$ label the spin
 positions on a suitable lattice of spatial dimension $d$,
and $r_{ij}$ is the inter-spin spatial separation.
Each spin satisfies $|\vec S|= 1 $ and   $0 \leq \alpha  < \infty $ 
parametrizes the range of interactions:
decreasing range for increasing $\alpha$, so that $\alpha=0$
corresponds to an all-to-all interacting model (close
to phenomenological anisotropic models), while $\alpha=\infty$
refers to a nearest neighbor interacting spin model.
$-1< \eta \le 1 $ parametrizes the degree of anisotropy and for 
$\eta=-1$ the Hamiltonian (\ref{H}) does not have a single 
easy axis.

In Eq.~(\ref{H}) the constant $J>0$ has been added in order to fix the scale
of time and to describe the model as ferromagnetic.

Needless to say, this is not
the most general spin Hamiltonian 
giving rise to a TNT, even if it can be considered the simplest
non integrable Hamiltonian with a suitable
energy threshold above which the  phase space is divided.

The minimum energy configuration,  with energy $E_{min}$, is attained when 
all spins are 
aligned along the $Z-$axis\cite{tutti} which defines implicitly 
the easy axis of magnetization.

The phase space for $E=E_{min}$ contains only two spin configurations,
parallel or anti-parallel to the $Z-$axis.  
Therefore, the phase space at the minimal
energy is disconnected, due to the uniaxial anisotropy and 
it consists of two points only.
We may ask now when and whether  at  a higher energy the
constant energy surface is connected.    
To this purpose, let us  define the TNT energy $E_{tnt}$ as 
the minimum energy compatible with the constraint of 
zero magnetization along the easy axis of magnetization
(hereafter we call $m$ the magnetization along the easy axis)
$$
E_{tnt} = {\rm Min } \ \left[ \ H  \ \left| \ m \equiv  \sum_i S_i^z/N 
=  0 \ \right]   \right..
$$
It is clear that if   $E_{tnt} > E_{min}$   the phase space
will be disconnected for all energies  $E < E_{tnt}$.
We call this situation Topological Non-connection, and, as will become clear in a moment,
its physical (dynamical as well as statistical) consequences  
are rather interesting. 
Indeed, since below the TNT the phase space is disconnected,
no energy conserving dynamics can bring the system from a configuration with 
$m>0$ to a configuration  with $m<0$, thus indicating
an ergodicity breaking
(impossibility to visit the whole energy surface). 

A useful quantity measuring how large
the disconnected energy region is, 
compared to the total energy range, 
can be
introduced\cite{tutti}:
\begin{equation}
\label{cxy81}
r  = \frac{E_{tnt}-E_{min}}{|E_{min}|}. 
\end{equation}
In \cite{tutti} it has also been  shown that the
disconnection ratio $r$, for  $N\to \infty $,
\begin{equation}
\label{cxy}
r    \to \left\{ \begin{array}{lll}0 
&&  {\rm for } \qquad \alpha \ge d  \\const \ne 0  
&&     {\rm for } \qquad \alpha < d,
\end{array}\right.
\end{equation}
where $d$ is the dimension of the embedding lattice.
Since this point has not been remarked in Ref.~\cite{tutti}, let us stress
here that the existence of a phase transition for $\alpha < d $
can be inferred from the finiteness of $r$
in the thermodynamic limit. 
Indeed, for long range systems,
in order to define the thermodynamic limit it is  convenient 
to make the energy extensive.  This can 
be achieved by multiplying the Hamiltonian by $N/|E_{min}|$.  
If we  define  the energy per 
particle, $e=E/N$, we can write:
$$
e_{tnt}-e_{min}=\frac{N}{|E_{min}|} \left( \frac{E_{tnt}}{N}-\frac{E_{min}}{N}
\right) \equiv r.
$$
Since below  $e_{tnt}$ the most probable magnetization 
is for sure different from 
zero, then the specific energy at which the most probable magnetization 
is zero will be 
greater than the minimum energy in the thermodynamic limit, 
thus implying a phase transition. On the other hand, 
let us also notice that when 
$r \to 0 $  neither the existence nor the
absence of a phase transition can be deduced.

An estimate for the TNT was also given for $\alpha < d $ and
large $N$. More precisely,  it can be shown that\cite{tutti}:
\begin{equation}
\label{cxy1}
E_{tnt}  \approx \left\{ 
\begin{array}{lll} 
4E_{min}^\prime -E_{min} &&  {\rm for } \qquad \eta \ge \eta_{cr} \\ 
&&\\
-\eta E_{min}   &&     {\rm for } \qquad \eta < \eta_{cr}, 
\end{array}\right.
\end{equation}
where $$\eta_{cr} \simeq 1-2^{\alpha/d} <0,$$ and $E_{min}^\prime$ is
the minimal energy for a system of $N/2$ spins. 
For $\eta> \eta_{cr}$ the TNT is given by the minimum of the second term
in Eq.~(\ref{H}) under the constraint $m=0$, while for $\eta < \eta_{cr}$
it is given by the minimum, under the same constraint, of the first term
in Eq.~(\ref{H}).
For $\eta<0$ and finite $N$,
there is a competition between the two different TNTs, therefore, 
in  what follows, we will fix the anisotropy parameter $\eta=1$.
Needless to say this choice does not affect the generality of our results.

Due to the disconnection, below 
 $E_{tnt}$ the dynamical 
time of magnetic reversal 
is infinite 
while above and close to the energy threshold
(for chaotic systems), it was found to diverge as a power law\cite{firenze}:

\begin{equation}
\tau \sim \frac{1}{(E-E_{tnt})^{\gamma}} \sim 
e^{\Delta S} = \frac{P_{max}}{P_0},
\label{taum}
\end{equation}
where $\Delta S$ is the entropic barrier between the most likely magnetic 
states.
Here,  $P_E(m)$ is the probability distribution of the magnetization $m$
at fixed 
energy $E$, so that  
$$P_{max} \ = \ \displaystyle{\rm Max}_m \  P_E(m),$$
 and $P_0=P_E(m=0).$

The divergence found in \cite{firenze} also shows that the phase 
space becomes connected 
at $E_{tnt}$, a non trivial result, which cannot be  deduced from 
the true existence of $E_{tnt}$.

Another important result found in \cite{firenze} 
 is that, for all-to-all
interacting spins ($\alpha = 0$),  the exponent 
$\gamma \propto N$.
This is related with  the extensivity of the entropy 
$S(E,m) = \ln P_E(m)$  (here we set  $k_B = 1 $) and explain
the  huge metastability of such states even for
small systems (say $ N \sim 100$) 
and  not necessarily close to the threshold $E_{tnt}$.

We numerically checked that, even for other power-law decreasing potentials
in the long-range case $\alpha < d$, 

\begin{itemize}
\item
a power law divergence at $E_{tnt}$ still occurs
as given by Eq. $ \ $(\ref{taum});
\item reversal time is still 
proportional to $P_{max}/P_0$ (same equation);
\item the exponent
$\gamma = N$.
\end{itemize}
 
In order to do that 
we computed $P_E( m=0 )$ and $P_{max}$ for different systems 
using the Wang-Landau algorithm \cite{wanglandau}.

\begin{figure}[ht]
\includegraphics[scale=0.35, angle=-90]{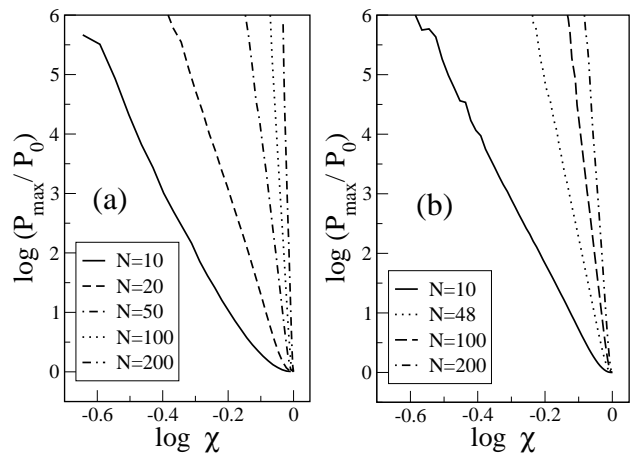}
\caption{
$\log P_{max}/P_0$  {\it vs} $\log \chi $
for different $N$  values as indicated in the legend and 
(a)  $\alpha=0.5$; (b)  $\alpha=1$.
}
\label{fa0}
\end{figure}

In Fig. \ref{fa0} we show the power law divergence 
of $P_{max}/P_{0}$ for different $\alpha$ and $N$ values.
In order to improve the presentation 
we choose  as a variable on the  $X-$axis 
\begin{equation}
\label{chi77}
\chi = \frac{E-E_{tnt}}{E_{stat}-E_{tnt}},
\end{equation}
where $E_{stat}$ has been
defined as the energy at which $P_{max}= P_0$ (that is when 
the probability distribution of the magnetization has
a single peak). 
That way all curves have a common origin.

\begin{figure}[ht]
\includegraphics[scale=0.35, angle=-90]{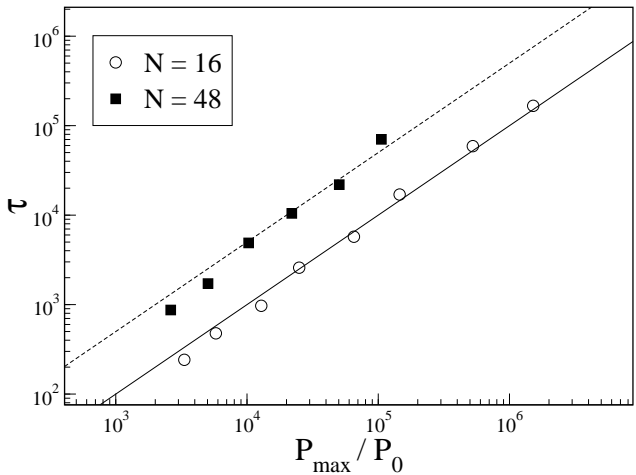} 
\caption{Dynamical reversal time $\tau$ {\it vs}
the statistical one $P_{max}/P_0$ for $\alpha=1$
and different $N$ values as indicated in the legend. 
Straight lines are  
$\tau = (1/k) P_{max}/P_0$ with $k=10$ for $N=16$ and $k=2$
for $N=48$. They have been drawn with the only purpose to guide 
the eye showing the proportionality between the two quantities over 3 orders
of magnitude.
}
\label{fa1}
\end{figure}

When the isolated system has a chaotic dynamics we computed the
magnetic reversal time from the direct integration of the 
equations of motion and we compare it with the "statistical"
time $P_{max}/P_0$ as given by Eq.~(\ref{taum}).
We show this comparison in Fig. \ref{fa1} where each point
on the graph has  a $X$ coordinate $P_{max}/P_0$ and
a $Y$ coordinate $\tau$.
The straight lines indicated proportionality over 3 orders 
of magnitude.

The linear dependence $\gamma = N$
found in \cite{firenze} for the case $\alpha=0$ holds 
for generic $\alpha$ too.

\begin{figure}[ht]
\includegraphics[scale=0.34,angle=-90]{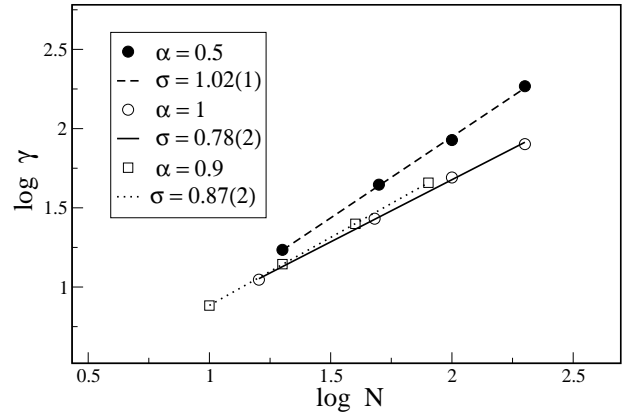}
\caption
{$\log \gamma$ as a function of $\log N$ for
different $\alpha$ values. Full circles stand for $\alpha=0.5$ 
and dashed line is the linear fitting with slope $1.02(1)$.
Open circles is the critical case $\alpha=d=1$ and full line
is the linear fit with slope $0.78(2)$. 
Open squares are for $\alpha=0.9$. Standard fit procedure gives 
$\sigma=0.87$ thus signaling the presence of a smooth transition
at the point  $\alpha= 1$ for finite $N$.
}
\label{fnew3}
\end{figure}

In Fig.~\ref{fnew3} we show
the results of our numerical simulation
for  $\alpha=0.5, 0.9, 1$ . Each point $\gamma$, at fixed $N$ 
has been obtained  computing the statistical
reversal time  for different  energies, as plot in Fig. \ref{fa0}, 
using the power law  (\ref{taum}).
Assuming  a power law dependence  $\gamma \propto N^\sigma$ 
we have found $\sigma \approx 1$ (within numerical errors) 
for all cases $\alpha \ll  1$
(we show for simplicity only the case $\alpha=0.5$
in Fig. \ref{fnew3}).

On the other hand, for $\alpha \sim 1$ we have numerical evidence
of a slower dependence on $N$: $\gamma \sim N^\sigma$ with $\sigma < 1$.
In the same 
Fig.~\ref{fnew3} we show for sake of comparison the critical case $\alpha=d=1$
where 
$\sigma =0.78(2)$, and 
the  close-to-critical case $\alpha=0.9$ where we have found $\sigma=0.87(2)$.
Even if these results 
indicate that the simple linear  
relation $\gamma \propto N$
is valid for long-range interacting systems only,
care should be used to extend the results of the case $\gamma=d=1$ 
to large $N$ since finite $N$ effects are huge in this case. 
Numerical evidence for $\sigma < 1$ has also been found in the short
range case ($\alpha \gg 1 $ ) but  it will be discussed 
elsewhere.

\section{Density of States}

The density of states (DOS) for a  Mean-Field approximated
Hamiltonian can be computed analytically,
using large deviations techniques \cite{Jul}.
In particular we will show that $\rho(E) \simeq (E-E_{min})^N$
for $E$ close to $E_{min}$. We  will also give numerical evidence 
that this law still constitutes 
an excellent approximation of the full Hamiltonian (\ref{H})
and for generic power law interaction  $\alpha \ne 0$.

Let us consider the following Mean-Field Hamiltonian:

\begin{equation}
H_{mf}=-\displaystyle \frac{J}{N} \left(\sum_k S^z_k\right)^2,
\label{MF}
\end{equation}
which can be considered a Mean-Field approximation of the Hamiltonian 
(\ref{H}), for low energy and $\alpha=0$.
Defining $$m_z=\frac{1}{N} \sum_k  S^z_k$$ and $e=H_{mf}/N$, 
Eq.~(\ref{MF}) can be rewritten as  
$$e=- J m_z^2.$$

Let us also assume that $S^z_k$ are random variables uniformly distributed
in $[-1,1]$.

We can compute the entropy per particle as a function of $m_z$
using Cramer's theorem \cite{Jul}, so that we have:

\begin{equation}
s(m_z)= - sup_{\lambda} \left[ \lambda m_z - \ln {\psi( \lambda)} \right],
\label{s}
\end{equation}
where 
$$
\psi(\lambda)= \langle e^{\lambda S^z}\rangle= \frac{1}{2}
 \int_{-1}^{1} e^{\lambda S^z} d S^z= \frac{e^{\lambda}-e^{-\lambda}}{2 \lambda}.
$$
Taking  the $sup$ in Eq.~(\ref{s})   we get:

\begin{equation}
\frac{\psi^{'}(\lambda)}{\psi(\lambda)}=m_z,
\label{sup}
\end{equation}
which defines $\lambda$ as a function of $m_z$.
It is easy to see that for $m_z \sim 1$ then $\lambda\to\infty$
(we could as well consider $m_z \sim -1$ of course, and the result
would be the same) so we restict our
considerations to 
$|m_z|\simeq 1$.
Simplifying  the expression of $\psi$  we have 
$$
\psi(\lambda) =\frac{e^\lambda}{2\lambda},
$$
and inverting  Eq.~(\ref{sup}):

$$
\lambda=\frac{1}{1-m_z}\equiv \frac{1}{\delta}.
$$

\begin{figure}[ht]
\includegraphics[scale=0.35,angle=-90]{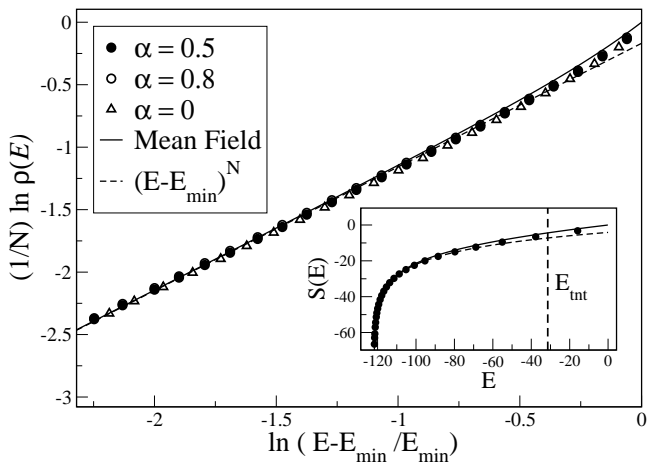}
\caption
{ The specific entropy {\it vs}  energy,  obtained numerically 
for $N=24$ and 
different $\alpha$ values (symbols indicated in the legend) 
is
compared with that of the Mean Field Hamiltonian (full curve) 
and with the power law (dashed line), see Eq.~(\ref{DOSpl}). In the inset
the entropy is shown {\it vs} the energy for the case $N=24$, $\alpha=0.5$. }
\label{fnew4}
\end{figure}

From Eq.~(\ref{s}) we obtain:
\begin{equation}
\label{sebis}
s(\delta)= \ln{\delta} + const. =
\ln\left( \frac{e-e_{min}}{-e_{min}} \right) + const.,
\end{equation}
since 
$$
e=- J m_z^2 \simeq -J(1 - 2 \delta)= e_{min} + 2 \delta J,
$$
and $e_{min}=-J$. From that we immediately have that at low energy, 

\begin{equation}
\rho(E) \simeq (E-E_{min})^N .
\label{DOSpl}
\end{equation}
The next leading order in $\delta$ can be calculated 
from 
$$
e=-Jm_z^2 =-J(1-\delta)^2,
$$
so that 
$$
\delta=1-\sqrt{-\frac{e}{J}},
$$
and  
\begin{equation}
s(e)= \ln \left(1-\sqrt{\frac{e}{e_{min}}} \right) + const.
\label{se2}
\end{equation}
It is immediate to see that Eq.~(\ref{se2}) gives Eq.~(\ref{sebis})
for $e \simeq e_{min}$.
We compared this analytical result for the Mean-Field model 
with the DOS  computed numerically for the 
full Hamiltonian (\ref{H}) and different $\alpha$ values
in Fig. \ref{fnew4}.
The DOS has been calculated using a modified Metropolis algorithm
introduced in \cite{hove}. The idea behind 
 is  performing  a random walk 
in phase space within an energy range defined by the system 
temperature. The probability $P(E,T)$
of visiting a configuration with energy $E$ and  temperature $T$, 
obtained 
keeping a histogram of the energy values found  during a Metropolis run, is
related to the DOS through the Boltzman factor 
$\exp(-E/T)$. That
 provides us  a conceptually simple way of determining the DOS.
However, due to finite run time, $P(E,T)$ will only contain information near
$\langle E\rangle $  and we must combine the results for runs 
at different temperatures
to obtain the complete DOS over the entire energy range.

As one can see  in Fig. \ref{fnew4}
the entropy
per particle, in the long range case, is almost
independent of  the range of the interaction, also
confirming a  result 
obtained in \cite{toral}. Moreover, the  theoretical
approach gives a very good approximation of the entropy 
per particle at low  energy. 
When the energy is increased, the first term in the full
Hamiltonian (\ref{H}) becomes important and some
deviations appear (see for instance the upper right corner
in Fig. \ref{fnew4}). Needless to say the excellent agreement 
 confirms the power law behavior for the DOS,
Eq.~(\ref{DOSpl}),  even for energy values sufficiently high. 
As an example in the inset of Fig.~\ref{fnew4}, we can 
see that the power law expression  (\ref{DOSpl}) holds up to $E_{tnt}$.

\section{magnetic reversal time}

Since the TNT has been introduced for isolated systems, question
arises if and how it can be defined  when  
the system is in contact with a thermal bath.
From the theoretical point of view we might expect 
that due to thermal noise the magnetization
will be able, soon or later,  
to change its sign at any temperature $T$, thus
suppressing the ergodicity breaking found in  isolated systems.
Therefore, strictly speaking,  
 a critical temperature below which the phase space 
is topologically disconnected for open finite systems does not  exist.

Nevertheless we are here interested in  more practical
questions, for instance:
Will the energy threshold $E_{tnt}$ 
still determine the magnetic reversal time
in presence of temperature 
as it does in isolated  chaotic systems? 
Can we predict the dependence of reversal time from
temperature or any other system parameters, like the number
of particles?

Since the system is in contact with a thermal bath we may
consider it as a member of a canonical ensemble. We may
properly define the probability density to have a certain
energy value $E$ at the temperature $T$: $P(E, T)$.
Considering all members of the ensemble as independent 
objects we may guess that when the average energy $\langle E \rangle$
is much less than $E_{tnt}$ and the probability density
$P(E,T)$ sufficiently peaked around its average value, the majority
of the members of the ensemble will not cross the barrier,
or at least, the probability of crossing it
will be small.
On the other hand for mean energy $\langle E \rangle$  on the order
of   $E_{tnt} $ each member  will be allowed to jump, with a time 
essentially given by the  microcanonical expression Eq.~(\ref{taum}).

Let us   further assume, following the standard fluctuation theory
\cite{griffith, landau}, that the magnetic reversal times 
between states with opposite magnetization are determined by the free 
energy barrier $\Delta A$ between  states at the most probable magnetization
and states with zero magnetization:

 \begin{equation}
\tau \propto \exp \left( \frac{\Delta A}{T}\right) =
 \frac{ {\rm Max}_m [P_T(m)] }{P_T (m=0)},  
\label{grifeq}
\end{equation}
where $$P_T(m) = \exp  [-A(T,m)/T],$$ is the probability
density to have magnetization $m$ at the temperature $T$.
Since ${\rm Max}_{m}  [P_T( m )] $ is usually a slow varying function of the
temperature, we can write 
$$\tau \propto \frac{1}{P_T (m =0)}.$$
The crucial point now, is to obtain such value using the microcanonical
results obtained in the previous Section, namely:

\begin{equation}
P_T (m =0)=\frac{1}{Z_T} \int P_E(m=0) \ e^{-E/T} \ \rho(E) \ dE,
\label{p0t}
\end{equation}
where $\rho(E)$ is the density of states and 
\begin{equation}
Z_T=\int \ e^{-E/T} \ \rho(E) \ dE,
\label{parfu}
\end{equation}
is the partition function.
Since $P_E(m=0)=0$ for $E<E_{tnt}$,
the ergodicity breaking acts as a cut-off energy of the integral
(\ref{p0t}), so that if the average energy is well
below $E_{tnt}$ 
we can expect very large  average reversal times.

In order to verify that Eq.~(\ref{p0t}) actually gives the 
magnetic reversal time, 
we simulated the dynamics of a spin system in contact with a thermal 
 bath in two different ways, 
the Metropolis algorithm \cite{metropolis}, and  using the stochastic differential
equations of the Langevin type as suggested in \cite{langevin}.

In the  Metropolis dynamics  the change in the spin direction has been taken 
at each step completely random on  the unit sphere while in the Langevin approach a small
dissipation has been added. We checked that the results are independent of such
dissipation and that the two approaches give the same results.

\begin{figure}[ht]
\includegraphics[scale=0.39,angle=-90]{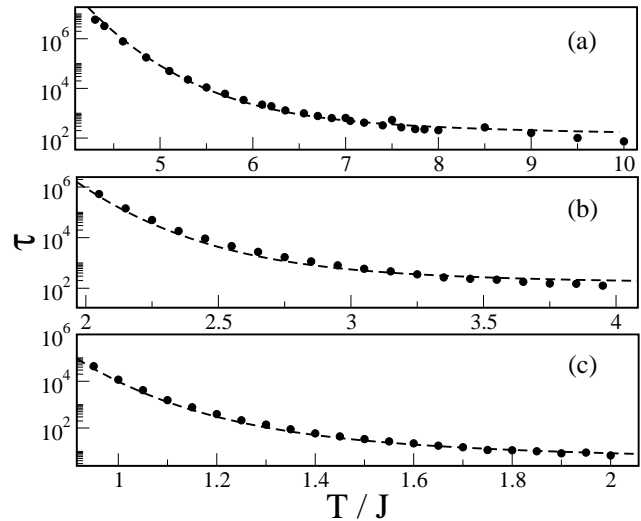}
\caption{Average reversal time $\tau$
as a function of the rescaled temperature $T/J$ for
different interaction range.
a) all-to-all $\alpha=0$, $N=24$;
b) long-range case $\alpha=0.5$, $N=24$;
c)  critical-case, $\alpha=1$,  $N=20$.
Circles  are numerical data, dashed line is the integral
calculated in Eq.~(\ref{p0t}).
}
\label{fau3}
\end{figure}

Obviously, both approaches 
can give directly the distribution function:
$P_T(m=0)$, but we prefer here the direct 
calculation of the density of states and thus
the possibility  
to get  $P_T(m =0)$ 
for  any temperature\cite{hove}, 
with less numerical effort and greater reliability.  
As for the reversal time it is quite obvious that an 
arbitrary multiplicative factor
should be considered when the two different approaches are compared
(the unit of time in the Metropolis approach is given by a random
spin flip).

Results are shown in Fig. \ref{fau3} where the average reversal time
(obtained with the Metropolis dynamics)  
{\it vs } the rescaled temperature $T/J$ has been considered 
for different $\alpha $ and $N$ values as indicated in the caption.
As one can see the agreement between the integral (\ref{p0t}) 
(dashed line in Fig. \ref{fau3}) 
and the numerical results (full circles) is excellent over 4 order of magnitude.
It is  worth of mention that no parameter fitting, other
than a multiplicative
constant has been used.

It is also remarkable that a small variation in the temperature scale (about a 
factor 2) generates a huge variation of the
average time (roughly 3-4 orders of magnitude).
This  signals an exponential dependence by the inverse temperature.
Nevertheless a simple temperature  dependence 
cannot be found in general, even if all curves 
in Fig.~\ref{fau3} can be fitted by the function
$T^a \exp (b/T)$, with $a$ and $b$ fitting parameters.

The exponential $1/T$ dependence can be understood in the 
limit of low temperature.  Unfortunately we cannot compute directly
the average time for  temperatures lower than those shown, due
to computer capability,  
even though we can study the asymptotic behavior of the integral (\ref{p0t})
for low temperature.

In order to obtain the low temperature behavior of the
magnetic reversal times, we can use the saddle point approximation to
Eq.~(\ref{p0t}) getting:
\begin{equation}
P_T (m =0) \simeq  P_{E^*} (m=0)\  e^{(\langle E \rangle - E^*)/T}   \
 e^{ S(E^*)  - S(\langle E\rangle)},
\label{p1t}
\end{equation}
where $\langle E\rangle$ and $E^*$ are given by:
\begin{equation}
\frac{1}{T} = \frac{dS}{dE}(\langle E\rangle) = 
\frac{ds}{de}(e),
\label{avet}
\end{equation}
where $S= \ln \rho$, 
and 
\begin{equation}
\frac{1}{T} = \frac{dS}{dE}(E^*) + \frac{N}{E^*-E_{tnt}} 
\label{estar}
\end{equation}
where we used $P_{E^*}(m=0) \propto (E^*-E_{tnt})^N$,
see Sect.~II. 
An approximate expression for (\ref{avet}) and (\ref{estar}) can be obtained 
for small temperature.
Indeed, using for the entropy the expression Eq.~(\ref{se2}) obtained in  
Sect.~III, and inverting    Eq.~(\ref{avet}),  one obtains:

\begin{equation}
\label{ev59}
\frac{e}{e_{min}} = \frac{1}{2} \left( 1+ \frac{T}{e_{min}} +
\sqrt{1+\frac{2T}{e_{min}} } \  \right),
\end{equation}
so that, for
$T \ll |e_{min}|/2$,
we get:

\begin{equation}
<E> = E_{min} \left[ 1 + \frac{NT}{E_{min}} +O\left(\frac{NT}{E_{min}}\right)^2
\right].
\label{avee}
\end{equation}
In the same way Eq.~(\ref{estar}) can be written as:
\begin{equation}
\label{estor}
E^* =  E_{tnt}+NT\left[1+
\displaystyle\frac{NT}{\Delta} + O\left(         
\frac{NT}{\Delta}
\right)^2 \right],
\end{equation}
where $\Delta=E_{tnt}-E_{min}$ and, 
for temperature sufficiently low, 
\begin{equation}
T \ll T_{cr} =\frac{E_{tnt}-E_{min}}{2N} < \frac{e_{min}}{2},
\label{tcr}
\end{equation}
we have  $$E^*=E_{tnt}+NT.$$

\noindent
Eq.~(\ref{p1t})  can be further simplified,
using the approximated expressions for $\langle E \rangle$
and $E^*$ obtained above  and Eq.~(\ref{DOSpl}) for the DOS: 

\begin{equation}
\label{arrn1}
\begin{array}{lll}
S(E^*) &\simeq & N \ \ln (\Delta +NT),\\
&&\\
S(\langle E\rangle) & \simeq & N \ \ln  NT,\\
&&\\
P_{E^*} (m=0) &\simeq &  (E^*-E_{tnt})^N \simeq (NT)^N.
\end{array}
\end{equation}

\noindent
Finally,  neglecting terms of  order  $ (NT/\Delta )$:

\begin{equation}
\tau  \sim \frac{1}{P_T(m=0)} \sim \exp\left(\frac{E_{tnt}- E_{min}}{T}\right).
\label{arr55}
\end{equation}
Even if  
Eq.~(\ref{arr55}) has been obtained for 
low temperature  
($T\ll T_{cr}$), it should be kept in mind
that this is a classical model so that
for $T \to 0 $, when quantum effects become important,
it looses its validity\cite{nota}).

\begin{figure}[ht]
\includegraphics[scale=0.35,angle=-90]{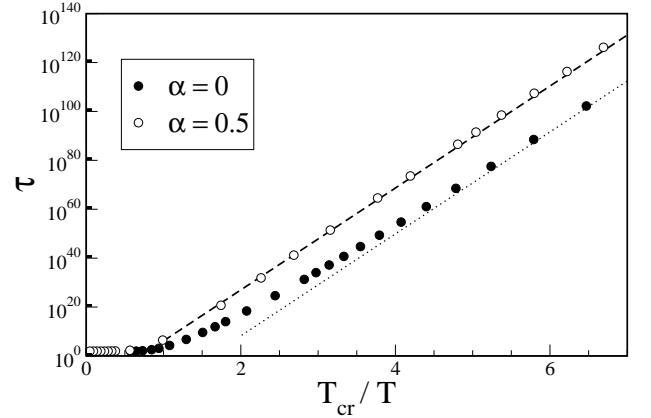}
\caption{Average reversal time $\tau$ calculated from 
Eq.~(\ref{p0t})
as a function of the rescaled temperature $T_{cr}/T$ for
different interaction range
$\alpha=0$ (full circles) and $\alpha=0.5$ (open circles)
and $N=24$. Dashed and dotted lines represent their 
asymptotic value, as given by Eq.~(\ref{arr55}).
}
\label{fau5}
\end{figure}

The law (\ref{arr55}) has been checked numerically in Fig. \ref{fau5}
where the integral (\ref{p0t}) has been calculated for very low
temperature and compared with  the true exponential law.
As one can see asymptotically they are very close over many order
of magnitude.

Eq. (\ref{arr55}) represents the  central result
of this paper. Indeed 
it allows to compute directly the reversal time
in presence of a thermal bath at low temperature $T$
without any complicated statistical calculations but
the knowledge of the Hamiltonian itself.
Moreover, the calculation of
both the ground state energy and the 
Topological Non-connectivity Threshold constitutes
a mechanical problem and they can be easily estimated even for 
complicated models.

Furthermore, it  also has some suggestive interpretation.
If we consider the path followed by the magnetization, 
as a random path of a Brownian particle between 
two potential wells separated by a potential barrier $\Delta U$,
according to Kramer's theory \cite{kramer, hanggi}
the average transition time between the two wells  follows
the Arrhenius law:

$$
\tau \sim \exp \left( \Delta U /T \right).
$$
Therefore it is clear that the disconnected energy region
$\Delta =  E_{tnt}-E_{min}$ can be thought of as the real potential barrier
felt by the magnetization.

In same way,  the critical temperature $T_{cr}$ has the physical meaning
of the specific energy barrier.
It is interesting to note that the condition $T \ll T_{cr}$
is not too restrictive, at least for long-range systems.
Indeed, taking into account that for large $N$ \cite{tutti}:
$$
T_{cr} = r \ \frac{|E_{min}|}{2N} \simeq \frac{2-2^\alpha}
{2(2-\alpha)(1-\alpha)} \ N^{1-\alpha},
$$
only at criticality ($\alpha = 1$) it does not depend on $N$
(and $T_{cr} = \ln 2$), while
generally it grows with the number of particles.
This is not at all surprising for a long range systems; indeed,
if we make the energy of the system extensive,
(multiplying the Hamiltonian by $N/|E_{min}|$),
we have  $T_{cr} = r /2$, which is finite for any interaction
range.

Last,  but not least, let us remark that
the model given by Hamiltonian ~(\ref{H}) at 
criticality ($\alpha=1$) is very interesting.
Indeed, for $\alpha=1$ the parameter  $r$ (the ratio between the 
disconnected portion of energy space compared to the full one)
goes to zero in 
the thermodynamical limit. The difference with the short range case
is that it goes to zero logarithmically, instead of a power law:
$r \sim 1/\ln N$.  This
simply means the existence of
an effective phase transition for finite systems
at criticality.

\section{Conclusions}

In conclusion we have shown that signatures of the topological
disconnection persist when a long range interacting spin system
is put in contact with a thermal
bath. 
More precisely for temperature sufficiently low we recover 
the Arrhenius law for the magnetization reversal time 
$\tau \propto \exp ( \Delta  /T )$ similar 
to the reversal time for a Brownian particle 
jumping across a potential barrier $\Delta  = E_{tnt}-E_{min}$.
In other words the magnetization behaves as a stochastic variable and
the potential barrier is exactly given by the energy distance
between $E_{tnt}$ and $E_{min}$. 
This proves  the exponential  dependence
of the reversal times from the number of particles 
and {\it stable} ferromagnetism even for small
systems with long range interaction and room  temperature.
The results presented in this paper can be experimentally verified,
using for instance the physical system discussed in \cite{expramaz},
or in a 3-D spin system with dipole interactions.


We acknowledge useful discussion with J.~Barr\'e and S.~Boettcher.
Financial support from PRIN 2005 and from  grant
0312510, DMR division of the NSF
is also acknowledged.


\begin{thebibliography}{}
\bibitem{nota} Classical spin models can be applied only 
when the temperature $T$ is sufficiently larger
than some characteristics quantum energy $\hbar \omega_0$.
\end{thebibliography}
\end{document}